\documentclass{aa}  

\usepackage{graphicx}
\usepackage{txfonts}
\usepackage{color}
\usepackage{bm}
\usepackage{media9}
\begin{document} 

   \title{Magnetohydrodynamic coronal modelling with high-fidelity magnetograms
   }
% \abstract{}{}{}{}{} 
% 5 {} token are mandatory
   \author{H. P. Wang \inst{1}
          \and
          S. Poedts\inst{1,2}
          \and
          W. S. Wang\inst{3}
          \and 
          R. Dhib\inst{1}
          \and
          A. Lani\inst{1}
          \and
          M. Condoluci\inst{1}
          \and
          T. Baratashvili\inst{1}
          \and
          Q. Noraz\inst{1,4}
          \and
          L. Linan\inst{1}
          \and
          H.-J. Jeong\inst{1,5}
          \and
          J. Y. Liu\inst{1,3}
          \and
          Z. G. Huang\inst{3}
          \and
          R. Liu\inst{3}
          \and
          L. Y. Dong\inst{1}
          \and
          R. Zhuo\inst{1}
          \and
          L. P. Yang\inst{6}
          \and
          G.  Pinz\'on \inst{7}
          \and
          J. Magdalenic\inst{1,8}
          \and
          B. Schmieder\inst{1,9,10}
          }
   \institute{Centre for Mathematical Plasma-Astrophysics, Department of Mathematics, KU Leuven, Celestijnenlaan 200B,
      3001 Leuven, Belgium\\
  \email{haopeng.wang1@kuleuven.be}\\
  \email{Stefaan.Poedts@kuleuven.be}\\
  \email{andrea.lani@kuleuven.be}
\and 
Institute of Physics, University of Maria Curie-Skłodowska, ul. Radziszewskiego 10, 20-031 Lublin, Poland 
\and
CAS Key Laboratory of Geospace Environment, Department of Geophysics and Planetary Sciences, University of Science and Technology of China, Hefei, 230026, People's Republic of China
\and
Rosseland Centre for Solar Physics, University of Oslo, PO Box 1029 Blindern, 0315 Oslo, Norway
\and
School of Space Research, Kyung Hee University, Yongin, 17104, Republic of Korea
\and
State Key Laboratory of Solar Activity and Space Weather, National Space Science Center, Chinese Academy of Sciences, Beijing, 100190, People's Republic of China
\and
Observatorio Astron\'omico Nacional, Universidad Nacional de Colombia, Bogot\'a, Colombia
\and
Solar-Terrestrial Centre of Excellence—SIDC, Royal Observatory of Belgium, 1180 Brussels, Belgium
\and
Observatoire de Paris, LIRA, UMR8254 (CNRS), F-92195 Meudon Principal Cedex, France
\and
LUNEX EMMESI COSPAR-PEX Eurospacehub, Kapteyn straat 1, Noordwijk 2201 BB, Netherlands}
  \abstract
  % context heading (optional)
  % {} leave it empty if necessary  
   { To resolve coronal structures in detail, it is essential to constrain the coronal model using high-fidelity magnetograms with minimal smoothing. } 
  % aims heading (mandatory)
    { In this paper, we improve the numerical stability of COCONUT to enable efficient coronal MHD simulations driven directly by photospheric magnetograms, including magnetic field strengths exceeding 1000 Gauss (G), and then use it to investigate the impact of small-scale magnetic field structures that are typically filtered out of photospheric magnetograms on the coronal simulation results.}
  % methods heading (mandatory)
   { Recently, a transition region module was implemented in COCONUT for solar-minimum coronal simulations constrained by magnetograms preprocessed by truncated spherical harmonics. In this work, we further adopt the recently developed energy decomposition strategy to enable it to remain numerically stable under solar-maximum conditions, allowing simulations driven directly by high-fidelity photospheric magnetic fields. We constrain this model with two types of magnetograms: the high-fidelity HMI photospheric magnetogram with only minor preprocessing through Gaussian smoothing, which preserves magnetic field strengths reaching 1500 G, and one processed using 50th-order filtered spherical harmonics, where the magnetic field strength is limited to below 100 G. We evaluate the impact of typically filtered small-scale magnetic field structures by comparing the differences between the two simulation results in terms of open magnetic flux (OMF), EUV images, polarised brightness (pB), and the squashing factor distributions.}
  % results heading (mandatory)
   { The results demonstrate that the COCONUT can reproduce the large-scale coronal structures revealed by pB and EUV observations. Compared with the coronal MHD simulation constrained by a conventionally filtered magnetogram, the simulation constrained by the high-fidelity magnetogram yields approximately $50\%$ more OMF and produces EUV images that are more consistent with the observations. Moreover, the results show that when viewed from low latitudes in certain EUV passbands, such as the 193~{\AA} channel, open-field regions near the solar poles may appear as bright structures that are typically associated with closed-field regions. This finding highlights the need for solar polar-orbiting missions to overcome the limitations and ambiguities arising from the restricted view of the polar regions from the ecliptic plane and to provide more complete full-Sun observations.}
  % conclusions heading (optional), leave it empty if necessary 
   {}

   \maketitle
%
%-------------------------------------------------------------------
\section{Introduction}
Due to the physical complexity and high computational cost of physics-based coronal  magnetohydrodynamic (MHD) modelling, practically applicable Sun-to-Earth modelling frameworks \cite[e.g.][]{Arge2010,Poedts_2020} typically rely on empirical coronal models that provide only coarse approximations of the coronal and solar-wind plasma conditions \citep{Arge2003ImprovedMF,Yangzicai2018,Merkin2016,Pomoell2018020,Shen_2018,Perri_2023}, thereby prioritising computational efficiency over physical fidelity \citep{Samara_2021}. Nevertheless, physics-based coronal MHD modelling \cite[e.g.][]{Brchnelova_2022,Feng_2012,FengandLiu2019,Feng_2021,Jin_2017,Hoeksema2020,Wang2022_CJG} is essential for accurately initialising downstream heliospheric modules and realistically simulating the propagation and evolution of solar disturbances \citep{Lionello_2013,van_der_Holst_2010,van_der_Holst_2022,TOTH2012870,Torok_2018,Poedts_2020,Baratashvili2024,Linan_2025}.
Considering that coronal MHD models capable of simultaneously achieving high computational efficiency, numerical stability, and physical fidelity remain scarce, the development of highly efficient, numerically stable, and spatio-temporally accurate coronal MHD modelling techniques is both urgent and essential. Such advances are crucial for deepening our understanding of the physical mechanisms underlying solar activity and space-weather events and, ultimately, for enabling reliable and comprehensive space-weather forecasting.

In recent decades, several coronal MHD models have been developed and implemented in advanced Sun-to-Earth model chains \cite[e.g.][]{Groth2000,Feng_2011Chinese,Feng_2013Chinese,Feng2020book,Jin_2012,TOTH2012870,Owens2017,Pomoell2018020,Hoeksema2020,Poedts_2020,Gombosi2021,Hayashi_2021,Yang_2021,Linker2024,Wraback_2025}. Among these coronal MHD models, those constrained by a static magnetogram can be classified as quasi-steady models, while those driven by time-evolving magnetograms can be referred to as time-evolving models. The quasi-steady-state coronal MHD models are generally used to reconstruct a static background coronal structure, whereas the time-evolving coronal MHD models are used to calculate the temporal evolutions of coronal structures. A more detailed description of these coronal MHD models is available in \cite{Feng2020book}, \cite{Wang2025_FirsttimeevolvingCOCONUT,wang2025COCONUTMayEvent,wang2025sipifvmtimeevolvingcoronalmodel,Wang_COCONUT_DecE} and \cite{Downs2025PSI}. Most of these models employ explicit or semi-implicit temporal integration schemes, in which the time step is constrained by the Courant–Friedrichs–Lewy (CFL) stability condition to only a few seconds or less. This severely limits computational efficiency when simulating the continuous evolution of the coronal structures over several weeks or longer.
To improve computational efficiency and numerical stability, a series of novel numerical algorithms have been developed and implemented in time-evolving coronal MHD models. Representative examples include the COolfluid COroNal UnsTructured model \citep[COCONUT;][]{Wang2025_FirsttimeevolvingCOCONUT,wang2025COCONUTMayEvent,Wang_COCONUT_DecE,wang2026OpenFlux_ApJ}, which is built on the Computational Object-Oriented Libraries for Fluid Dynamics (COOLFluiD) framework \citep{lani1,kimpe2,lani13}\footnote{\url{https://github.com/andrealani/COOLFluiD/wiki}}, and the Solar Interplanetary Phenomena Implicit Finite Volume Method \citep[SIP-IFVM;][]{wang2025sipifvmtimeevolvingcoronalmodel}, which employs a six-component composite grid \citep{Feng_2010,Feng2020book,Wang_2022,WangSIPtheoriticalCME}.

In COCONUT and SIP-IFVM, fully implicit time-integration schemes enable flexibility in selecting large time steps that substantially exceed the CFL stability constraint imposed on explicit methods, thereby significantly improving computational efficiency. Meanwhile, the reduction in temporal accuracy associated with the large time steps is alleviated via pseudo-time marching methods and inner iterations performed within each physical time step \citep{wang2025sipifvmobservationbasedmagnetohydrodynamicmodel,WangSIPtheoriticalCME,Wang2025_FirsttimeevolvingCOCONUT}. Additionally, the numerical stability of COCONUT is obviously improved by the positivity-preserving methods implemented in density, pressure, and temperature \citep{Wang2025_FirsttimeevolvingCOCONUT,wang2025COCONUTMayEvent,wang2026coconutpracticaltimeevolvingsuntoearth}, and especially by the decomposed energy strategy combined with an Harten–Lax–van Leer (HLL) Riemann solver that adds an extra dissipation term in the energy equation \citep{Wang_COCONUT_DecE}. Moreover, the numerical stability of SIP-IFVM is improved by the extended magnetic field decomposition method \citep{wang2025sipifvmtimeevolvingcoronalmodel}, which splits the magnetic field $\mathbf{B}$ into a static component $\mathbf{B}_{00}$, a temporally piecewise-constant component $\mathbf{B}_{01}$, and a time-evolving component $\mathbf{B}_1$. It maintains $|\mathbf{B}_1|$, which is updated at each time step, consistently small throughout the coronal evolutions. Both models are driven by heavily smoothed photospheric magnetograms, in which the small-scale magnetic field structures have been filtered out to extrapolate the photospheric magnetic field into the low corona, thereby reducing the maximum inner-boundary magnetic field strength to not significantly exceed 100 G. This oversimplified extrapolation does not account for the dynamical evolution governed by the MHD equations and may therefore lose critical physical information associated with the fine-scale structures evident in the photospheric observations.

To enable coronal simulations to be driven directly by photospheric magnetograms, it is necessary to incorporate chromospheric and transition-region (TR) modules to provide a bridge between the photospheric magnetograms and the low corona.
In the Alfv{\'e}n Wave Solar atmosphere Model (AWSoM) and Magnetohydrodynamic Algorithm outside a Sphere (MAS) coronal models, the transition region is artificially broadened using a specialised solution-preserving technique \citep{Lionello_2009,Mikic_2013,MIKIC2018NatA,Mok_2005}. This approach broadens the high-density plateau region by increasing the chromospheric density and prescribing a constant thermal conductivity below a reference temperature, while limiting the radial cell size near the solar surface to no more than 1~Mm to preserve the required spatial resolution \citep{Mok_2005,Lionello_2009,Mikic_2013,MIKIC2018NatA,van_der_Holst_2022}. To improve computational efficiency and numerical stability, the updated AWSoM model, AWSoM-R (AWSoM-Realtime), solves the one-dimensional equations governing plasma motion and heat transport along fixed potential magnetic field lines, together with Alfv{\'e}nic wave propagation in the low corona, and couples this threaded-field-line model to the full-MHD global coronal model at $1.1\;R_s$  \citep{Sokolov2021}. 
Alternatively, MAS \citep{Lionello_2023,Mason_2023} solves the resistive, thermodynamic MHD equations using a semi-implicit approach, in which a dispersive term is introduced to ensure numerical stability, allowing time steps to be defined by the CFL constraint determined primarily by the coronal flow speeds \citep{Lionello1999346}.

Recently, a simplified TR module, in which the inner-boundary density and temperature are specified to approximate the low chromosphere, was implemented in COCONUT and validated for a solar-minimum case (Tinatin et al. 2026), demonstrating its potential to provide a more realistic solar atmosphere. In this paper, we further incorporate this module into the energy-decomposed version of COCONUT \citep{Wang_COCONUT_DecE}, which has demonstrated sufficient numerical stability for a wide range of coronal simulations, including those under solar-maximum conditions. 
However, the TR module introduces a steep temperature gradient near the solar surface, resulting in a significantly increased thermal conduction term. 
We therefore also account for the saturated heat flux, which is limited by the local sound speed \citep{Cowie1977ApJ,Balbus1986ApJ,Meyer2012,Xia_2018}, when evaluating the thermal-conduction term. Together, these approaches enable efficient coronal simulations driven directly by photospheric magnetograms, which help to retain the fine-scale magnetic structures.

Based on the above considerations, this paper is organised as follows. In Section \ref{NumericalAlgorithm}, we describe the numerical algorithm, grid mesh, and preprocessing of the photospheric magnetograms used for the coronal simulations. In Section~\ref{sec:Numerical Results}, we present simulation results for the selected Carrington rotation, CR~2124. 
We compare coronal simulation results constrained by high-fidelity and filtered magnetograms, with the small-scale magnetic structures in the latter removed through a potential-field reconstruction based on truncated spherical harmonics.
Differences in the open magnetic flux, EUV emission, polarized brightness (pB), and squashing factor are evaluated to quantify the impact of the small-scale magnetic structures that are commonly excluded from coronal MHD simulations.
In Section~\ref{sec:Conclusion}, we summarise the features of the MHD modelling approach constrained by high-fidelity magnetograms and provide concluding remarks. 

\section{Numerical algorithm}\label{NumericalAlgorithm}
The quasi-steady-state coronal simulations in this paper are performed in a co-rotating Carrington heliographic coordinate system \citep{Burlaga1984MHDPI,FRANZ2002217}. The governing equations, spatial discretisation, and temporal integration methods are basically the same as those described in \cite{Wang_COCONUT_DecE} and \cite{wang2026coconutpracticaltimeevolvingsuntoearth}, with the main differences lying in the definitions of the thermal conduction, heating, and radiative loss terms. 

\subsection{The governing equations}\label{Governingequations} 
The governing equation adopted in this paper can be described in the following form:
\begin{equation}\label{MHDinsolarwind}
\frac{\partial \mathbf{U}}{\partial t}+\nabla \cdot \mathbf{F}\left(\mathbf{U}\right)=\mathbf{S}\left(\mathbf{U},\nabla \mathbf{U}\right) \,.
\end{equation}
In Eq.~\ref{MHDinsolarwind}, $t$ denotes the relaxation time over which the solution is evolved toward a steady-state coronal structure.  $\mathbf{U}=\left(\rho, \rho \mathbf{v}, \mathbf{B}, E_1, \psi\right)^T$ denotes the vector of conservative variables, $\nabla \mathbf{U}$ and $\mathbf{F}\left(\mathbf{U}\right)$ represent the spatial derivative of $\mathbf{U}$ and the inviscid flux vector. Here, $\rho$, $\mathbf{v}=(u,v,w)$, and $\mathbf{B}=(B_x,B_y,B_z)$ denote the mass density, velocity, and magnetic field, respectively, with the vector components expressed in Cartesian coordinates. $\psi$ is the Lagrange multiplier introduced in the hyperbolic generalised Lagrange multiplier method used for constraining the magnetic-field divergence error \citep{Dedner2002JCoPh,YALIM20116136}, $E_1=\frac{p}{\gamma-1}+\frac{1}{2}\rho\left|\mathbf{v}\right|^{2}$ refers to the decomposed energy density \citep{Wang_COCONUT_DecE}, $\gamma=\frac{5}{3}$ is the adiabatic index, and $p$ is the thermal pressure. Meanwhile, the gravitational force, the Coriolis force, and centrifugal acceleration, the source term caused by the decomposed energy
strategy \citep{Wang_COCONUT_DecE}, the empirically defined coronal heating $Q_{H}$ \citep{Mok_2005,Baratashvili2024}, the optically thin radiative loss term $Q_{rad}$ \citep{Wang2025_FirsttimeevolvingCOCONUT}, and the Spitzer or collisionless thermal conduction
term $-\nabla \cdot \mathbf{q}$ \citep{Hollweg1978,Mikic1999} are considered in the vectors of the source terms $\mathbf{S}\left(\mathbf{U},\nabla \mathbf{U}\right)$.  

Based on the previous experience \cite[e.g.][]{Mok_2005,Downs2010,Baratashvili2024,wang2025sipifvmobservationbasedmagnetohydrodynamicmodel,Wang2025_FirsttimeevolvingCOCONUT,Wang_COCONUT_DecE,wang2026OpenFlux_ApJ}, $Q_{H}$, $Q_{rad}$, and $\mathbf{q}$ can be defined in the following forms: 
\begin{equation}\label{HeatingSourceTerm}
Q_{H}=\eta \left(r\right) \cdot H_0 \cdot \left|\mathbf{B}\right| \cdot e^{-\frac{r-R_s}{\lambda}}\,,
\end{equation}
where $r$ and $R_s$ denote the heliocentric distance and solar radius, $\lambda = 0.7\,R_s$ and $H_0 = 8 \times 10^{-2}\,{\rm J\,m^{-3}\,s^{-1}\,T^{-1}}$ are adopted, and
\begin{equation}\label{eta_r}
\eta \left(r\right)=1+2\cdot \tanh\left(\frac{r-R_s}{R_s}\cdot100~\pi\right)
\end{equation}
is introduced to further adjust the empirically defined heating source term along the heliocentric distance \citep{wang2026OpenFlux_ApJ}.
The optically thin radiation loss term can be described as
\begin{equation}\label{Radiationloss}
Q_{rad}=-n_e \cdot n_p \cdot \Lambda\left(T\right)\,,
\end{equation}
where $n_p$ denotes the proton number density, which is assumed to be equal to the electric number density $n_e$ for the hydrogen plasma; $\Lambda\left(T\right)$ is a radiative cooling curve function \citep{Zhou2021,Wang2025_FirsttimeevolvingCOCONUT} derived from version 9 of the CHIANTI atomic database for emission lines \citep{Dere_2019}, and $T$ denotes the temperature. In chromosphere and TR, the solar atmosphere is considerably denser than the corona and is partially ionised.
We find that, near the solar surface, the current heating term is substantially weaker than the radiative loss term, rendering it insufficient to heat the corona to a temperature of million kelvin (K) and to accelerate the solar wind in the simulations considered in this paper.
Since this study mainly focuses on evaluating the impact of small-scale magnetic-field structures, which are commonly removed from photospheric magnetograms in coronal simulations, rather than on developing a self-consistent chromospheric and TR model, we further consider the following simply modified radiative loss prescription:
\begin{equation}\label{RadiationlossModified}
Q^{'}_{rad}=-n^{'}_e \cdot n^{'}_p \cdot \Lambda\left(T\right)\,,
\end{equation}
where $n^{'}_e=n^{'}_p$, $n^{'}_e=\left(1-\Upsilon\right)\cdot n_e+\Upsilon \cdot n_{\rm ref}$, and $\Upsilon=0.5+0.5\cdot\tanh\left(\frac{n_e-n_{\rm ref}}{\min\left(n_e,~n_{\rm ref}\right)} \cdot 10~\pi\right)$. Here $n_{\rm ref}=2\times10^8~{\rm cm^{-3}}$ is a reference number density defined to be comparable to the typical proton number density in the low corona.
In general, the heat flux is defined as
\begin{equation}\label{heatflux}
\mathbf{q}=\left\{\begin{array}{c}
-\xi T^{5 / 2}(\hat{\mathbf{b}} \cdot \nabla T) \hat{\mathbf{b}},  \text { if } 1 \leq r \leq 10~ R_{s} \\
\alpha n_{e} k_B T \mathbf{v},  \text { if } r>10~ R_{s}
\end{array}\right.,\
\end{equation}
where $\hat{\mathbf{b}}=\frac{\mathbf{B}}{\left|\mathbf{B}\right|}$, $\xi =9.0 \times 10^{-12}~\rm J~m^{-1}~s^{-1}~K^{-\frac{7}{2}}$, $\alpha=\frac{3}{2}$. For the convenience of description, we denote $\mathbf{q}$ as $\mathbf{q}_{\rm class
}$ hereafter.

The TR module introduces extremely steep temperature and density gradients near the solar surface, we therefore artificially broaden the transition region, following the approaches adopted in the AWSoM and MAS coronal models. Consequently, we calculate the heating and radiative-loss source terms as follows:
\begin{equation}\label{HeatingandRadiationSourceTerm}
Q^{*}_{rad}=Q^{'}_{rad}\cdot f\left(T\right)^{-1}\,,
\end{equation}
where 
\begin{equation}\label{BroadenFactor}
f\left(T\right)=\left(1+\left(\frac{T}{T_c}\right)^{-10}\right)^{\frac{1}{4}}  \,
\end{equation}
is a broadening factor (Tinatin et al. 2026) with $T_c=2\times10^5$ K. 

Considering that heat in a dilute plasma is transported primarily by electron motion, a larger thermal conduction term implies a higher electron thermal speed. However, when the electron speed becomes comparable to the sound speed, plasma instabilities will prevent the electrons from flowing much faster \citep{Cowie1977ApJ}. 
Therefore, in calculating the thermal conduction terms, we further account for the following saturated thermal flux $\mathbf{q}_{\rm sat}$ which is limited by the sound speed \citep{Cowie1977ApJ,Balbus1986ApJ,Meyer2012,Xia_2018}. 
\begin{equation}\label{heatfluxSat}
\mathbf{q}_{\rm sat} =\left\{\begin{array}{c} -5\cdot\phi\cdot\rho\cdot c_s^3\cdot{\rm sgn\left(\hat{\mathbf{b}} \cdot \nabla T\right)}~\hat{\mathbf{b}} ,  \text { if } 1 \leq r \leq 10~ R_{s} \\
-5\cdot\phi\cdot\rho\cdot c_s^3~\hat{\mathbf{v}},  \text { if } r>10~ R_{s}
\end{array}\right.,\
\end{equation}
where $\hat{\mathbf{v}}=\frac{\mathbf{v}}{\left|\mathbf{v}\right|}$, $c_s=\frac{\gamma\cdot p}{\rho}$ is the sound speed, $\phi$ is an uncertain number \citep{Balbus1986ApJ} which is set as $\phi=0.1$ in this paper.
Consequently, the thermal conduction flux in this paper is calculated as:
\begin{equation}\label{heatfluxLimitedbySat}
\mathbf{q}^{*}=\left(G\left(\frac{\left|\mathbf{q}_{\rm class}\right|}{\left|\mathbf{q}_{\rm sat}\right|}\right)\mathbf{q}_{\rm class}+\left(1-G\left(\frac{\left|\mathbf{q}_{\rm class}\right|}{\left|\mathbf{q}_{\rm sat}\right|}\right)\right)\mathbf{q}_{\rm sat}\right)\cdot f(T)\,,
\end{equation}
where the function limiter $G\left(x\right)$ is defined as following \citep{Meyer2012}:
\begin{equation}\label{FluxLimiter}
G\left(x\right)=\max\left(0,~\min\left(1,1-0.5x\right)\right)\,.
\end{equation}
These algorithms perform well for the background coronal simulations presented in this paper. However, since $\phi$ is an uncertain parameter in the definition of $\mathbf{q}_{\rm sat}$ \citep{Balbus1986ApJ}, the thermal conductive flux can still become excessively large under certain conditions, such as those associated with magnetic reconnection. We found that the algorithm can occasionally suffer from numerical stability issues in CME simulations. Therefore, when calculating thermal conduction, we further limit the temperature to $\min(T,T{\rm thre})$, where $T_{\rm thre}$ is an adjustable parameter. We adopt $T_{\rm thre}=1\times10^{6}$~K in this paper. This additional temperature limiter performs well for both the background coronal and CME simulations that we have tested.

\subsection{Boundary-condition setups and grid system}\label{BCinner}
In this paper, we synthesize a daily radial-field synoptic map from the HMI Polar-Filled synoptic maps \cite[hmi.synoptic\_mr\_polfil\_720s;][]{sun2018polarfieldcorrectionhmi}, which updates a synoptic map with the new observation centred at 13:00 UT on 2012 June 14, corresponding to CR 2124. It has a resolution of $1440\times3600$ uniformly distributed in $\sin \rm (latitude)$ and longitude.  To suppress small-scale fluctuations and discontinuities in the magnetic field distribution, we preprocess the synthesised magnetogram via the following Gaussian smoothing and then use the resulting magnetogram to prescribe the radial magnetic field at the inner boundary of the coronal model.
\begin{equation}\label{GaussianSmooth}
B_{r}^{*}(i,j)=
\sum_{m=-k}^{k}
\sum_{n=-k}^{k}
G(m,n)
B_{r}(i-m,j-n)\,,
\end{equation}
where $G(m,n)$ is the Gaussian kernel for a $\left(2k+1\right)\times\left(2k+1\right)$ kernel centred on magnetogram grid $\left(i,j\right)$, $B_r$ denotes the original radial photospheric magnetic field, and $B_{r}^{*}$ represents the smoothed magnetic field. In this paper, we set $k=4$, and define $G(m,n)$ as: 
\begin{equation}
G(m,n) =
\exp\left(-\frac{m^2+n^2}{2\sigma^2}\right)\cdot w\left(m,n\right)\,,
\end{equation}
where $\sigma=\frac{2k+1}{6}$, and $w(m,n) = \exp\left(-\frac{m^2+n^2}{2\sigma^2}\right)\cdot S_{i-m,j-n} \big/ \displaystyle\sum_{m=-k}^{k}\sum_{n=-k}^{k} \exp\left(-\frac{m^2+n^2}{2\sigma^2}\right)\cdot S_{i-m,j-n}$ with $S_{i-m,j-n}$ denoting the area of the $\left(i,j\right)$th magnetic pixel.
The magnetic field strength in this photospheric magnetogram reaches up to 1500 G. 
For comparison, we further filter the magnetogram using 50th-order filtered spherical harmonics \citep{MCCLARREN20105597,Murteira2025,wang2026OpenFlux_ApJ}, resulting in an over-filtered magnetogram in which the magnetic field strength is limited to 80 G. 

\begin{figure*}[htpb]
\begin{center}
  \vspace*{0.01\textwidth}
    \includegraphics[width=0.8\linewidth,trim=1 1 1 1, clip]{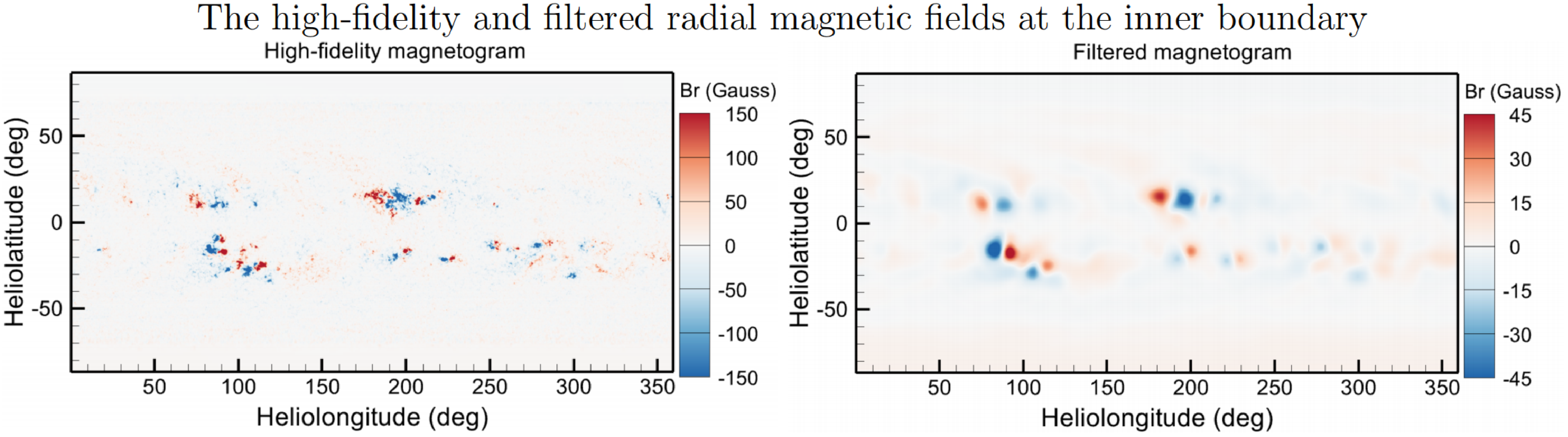}
\end{center}
\caption{Distributions of the high-fidelity (left) and filtered (right) radial magnetic field used to constrain the coronal simulations.}\label{InBDRadialMagneticField}
\end{figure*}

Figure~\ref{InBDRadialMagneticField} displays the distribution of the radial magnetic field in the two magnetograms, which are used to define the inner-boundary radial magnetic field of the coronal model, demonstrating that the high-fidelity magnetogram preserves substantially more small-scale magnetic field structures than the filtered magnetogram.
Meanwhile, the inner-boundary thermal pressure and plasma density are set to $1.1278~\rm Pa$ and $3.34 \times 10^{-9}~\rm kg~m^{-3}$, respectively, with the inner-boundary temperature being $2\times 10^4~\rm K$.  
Additionally, the density, the velocity vector, the tangential magnetic field at the inner boundary, and the outer boundary conditions are treated the same as in \cite{Wang2025_FirsttimeevolvingCOCONUT}.

\begin{figure*}[htpb]
\begin{center}
  \vspace*{0.01\textwidth}
    \includegraphics[width=0.8\linewidth,trim=1 1 1 1, clip]{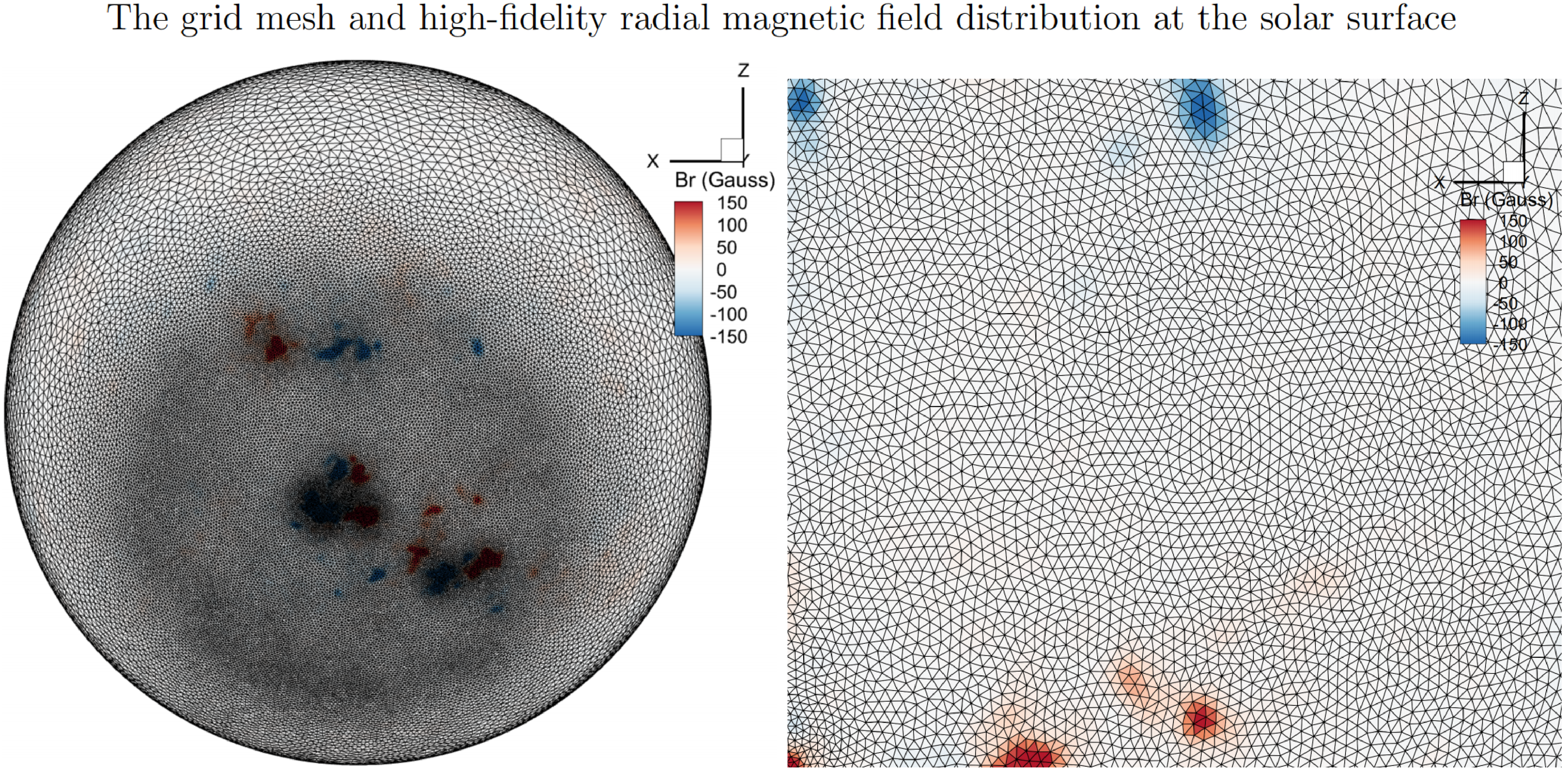}
\end{center}
\caption{The distribution of mesh nodes in each layer of the grid mesh, viewed along the $Y$ direction. The contours show the high-fidelity radial magnetic field distribution at the solar surface. The right panel shows a close-up of the left panel.}\label{GridMeshandRadialField}
\end{figure*}

The computational domain for the coronal simulation is a spherical shell extending from the solar surface to approximately 0.1~AU. There is an active region (AR) centred at $\left(\theta,~\phi\right)=\left(16.82^{\circ}~{\rm S},~ 87.85^{\circ}\right)$, which is approximately aligned with the Sun-Earth longitude \citep{WangWS2019}. To adequately resolve this AR, we locally refined the computational mesh over the region defined by $16.82^{\circ}~{\rm S}-45^{\circ}\le \theta \le16.82^{\circ}~{\rm S}+45^{\circ}$ and $87.85^{\circ}-45^{\circ}\le \phi \le 87.85^{\circ}+45^{\circ}$. The selected region is discretised using ninth-level subdivided unstructured geodesic meshes, with a smooth and continuous transition to sixth-level subdivided unstructured geodesic meshes \citep{Brchnelova2022} across the remainder of the computational domain. The sixth- and ninth-level meshes have tangential resolutions of approximately $1.78^{\circ}$ and $0.22^{\circ}$, respectively \citep{wang2025COCONUTMayEvent,Wang_COCONUT_DecE}. Figure~\ref{GridMeshandRadialField} presents the grid mesh distribution at a given heliocentric distance, together with the high-fidelity radial magnetic field distribution at the solar surface, viewed from the $Y$ direction. The close-up image (right) shows a gradual transition in mesh resolution from level-9 around the lower-left region to level-6 around the upper-right region.
In the radial direction, the computational domain consists of 164 layers of gradually stretched cells, with 70,112 cells in each layer. 
The first 80 layer of cells have a uniform radial width of $2\times10^{-4}~R_s$ (Tinatin et al. 2026), extending from the solar surface to 1.016~$R_s$. The remaining 84 layers gradually increase in radial width, reaching a maximum radial cell size of approximately 1~$R_s$ between heliocentric distances of 10 and 22~$R_s$.

\section{Numerical results}\label{sec:Numerical Results}
In this section, we assess the impact of small-scale magnetic structures, which are generally filtered out, on the coronal simulation results by comparing two coronal simulations constrained by the high-fidelity and filtered magnetograms. 
The two simulations are performed on 1920 CPUs of the wice cluster, part of the Tier-2 supercomputer at the Vlaams Supercomputer Centrum\footnote{\url{https://www.vscentrum.be/}}. The CFL number is gradually increased from 2 to 120 over the first 700 iterations and then maintained at 120 for the subsequent 2300 iterations. 
The CPU times required for the two coronal simulations with approximately 11.5 million prismatic grid cells, constrained by the high-fidelity and filtered magnetograms, are 1.37 and 1.28 hours, respectively. Hereafter, we use `MHD-HR' and `MHD-LR' to denote the coronal simulations constrained by the high-fidelity and filtered magnetograms, respectively

\subsection{Impact on the simulated EUV emission and open field}\label{sec:EUVandOMF}
Coronal holes are among the most prominent features of the solar corona, typically associated with open magnetic flux (OMF) regions and appearing as dark regions in EUV and soft X-ray images \citep{Petrie2011SoPh,FengMa2015,Feng_2017,wang2025sipifvmtimeevolvingcoronalmodel}. In this subsection, we present the EUV images and open-field regions at the solar surface obtained from MHD-HR and MHD-LR simulations. We also compare the OMF obtained from the two simulations.

\begin{figure*}[htpb]
\begin{center}
  \vspace*{0.01\textwidth}
    \includegraphics[width=0.8\linewidth,trim=1 1 1 1, clip]{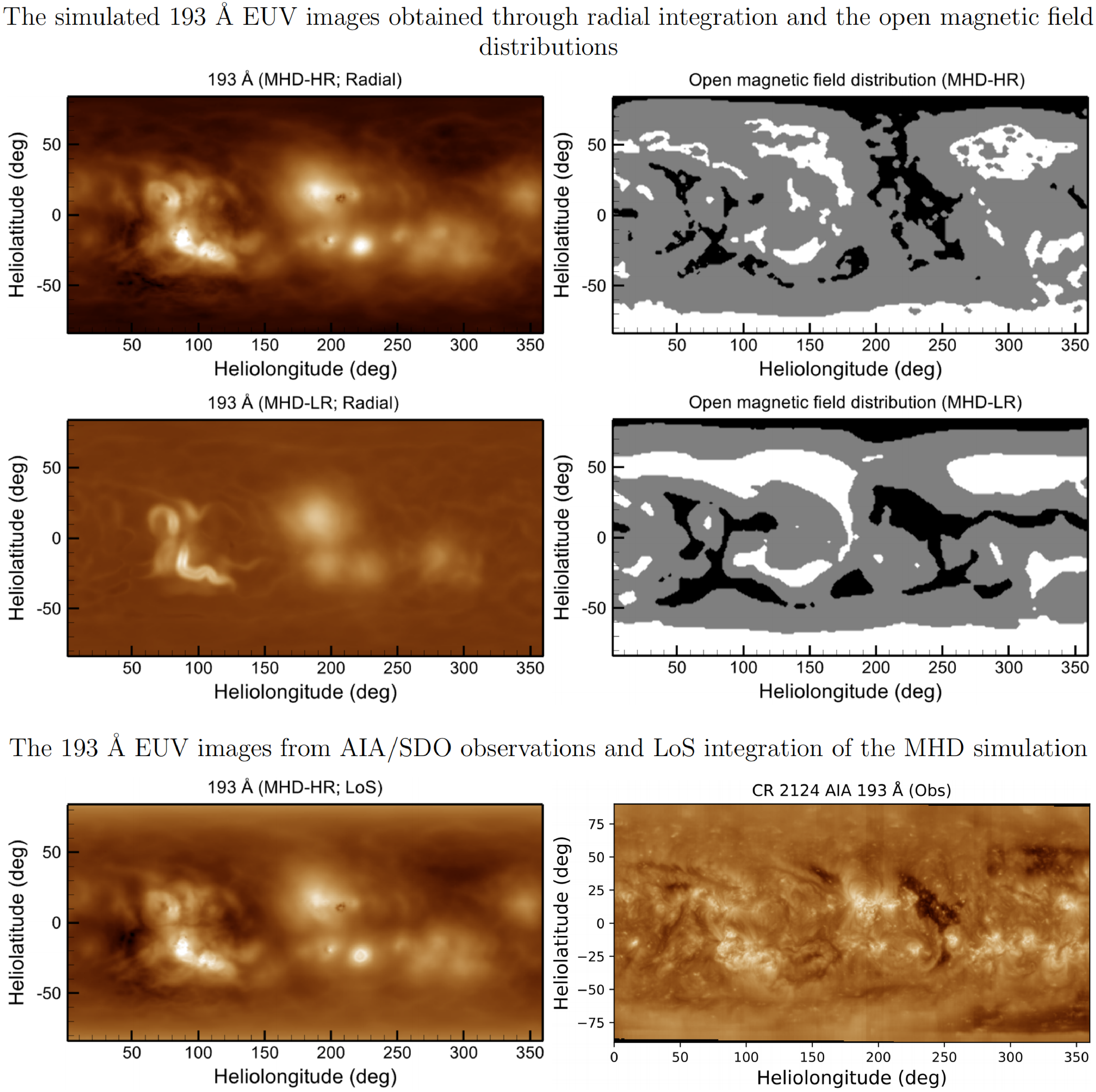}
\end{center}
\caption{Distributions of the 193~{\AA} EUV emission (left panels in all three rows and right panel in the third row) and simulated open-field regions (right panels in the first two rows). The white and black patches of the open-field region distributions represent open-field regions with magnetic field lines pointing outward and inward from the Sun, respectively, whereas the gray patches denote closed-field regions. The first and second rows denote the simulation results constrained by the high-fidelity (abbreviated as MHD-HR) and filtered (abbreviated as MHD-LR) magnetograms, respectively, and the third row displays the synoptic map of the 193~{\AA} EUV observation from AIA onboard SDO satellite.}\label{CHandEUV}
\end{figure*}

In Fig.~\ref{CHandEUV}, we present synthetic 193~{\AA} EUV images derived from MHD-HR (first and third rows) and MHD-LR (second row) simulations. The images in the first two rows are obtained by integrating the EUV emission radially from 3.5~$R_s$ to 1.002~$R_s$ \citep{wang2025sipifvmtimeevolvingcoronalmodel}, whereas the one in the third row is synthesised through line-of-sight (LoS) integration. For comparison, the lower-right panel shows a 193~{\AA} EUV image observed by the Atmospheric Imaging Assembly (AIA; \citealt{Lemen2012}) onboard the Solar Dynamics Observatory spacecraft\footnote{\url{https://sdo.gsfc.nasa.gov/data/synoptic/}} (SDO; \citealt{Pesnell_2012SoPh}). It indicates that the MHD-HR successfully reproduces both the bright structures presented in the 193~{\AA} EUV observations within $\pm40^{\circ}$ of the solar equator and the observed dark structures scattered around the bright structures. Whereas the MHD-LR simulation fails to reproduce the bright structure centred at $(\theta,\phi)=(20^{\circ}~\mathrm{N},345^{\circ})$. Moreover, the clustered, block-like bright structure around $(\theta,\phi)=(20^{\circ}~\mathrm{S},100^{\circ})$ and the clearly identifiable dark structures, both of which are evident in the observations and the MHD-HR simulation, are reduced to two simple ribbon-like bright structures and diffuse, poorly defined dark structures, respectively.

Figure~\ref{CHandEUV} also shows obvious differences between the polar-region EUV images obtained through LoS and radial integrations. To clarify the magnetic topology of these regions, we trace magnetic field lines from 15~$R_s$ to the solar surface to identify the open-field regions in the MHD simulations \citep{wang2026OpenFlux_ApJ}, as shown in the right panels of the first two rows. The results reveal that both polar regions are dominated by open magnetic fields. Nevertheless, when viewed from low latitudes, these regions appear bright in the LoS-integrated EUV images, although bright EUV structures are usually associated with closed-field regions. 
Additionally, the MHD-HR simulation successfully reproduces the open-field regions corresponding to the dark regions in the EUV observations, which are distributed within $\pm50^{\circ}$ of the solar equator and span longitudes from $0^{\circ}$ to $300^{\circ}$. The remaining discrepancies may be attributed to the lack of coronal evolution and transient solar activity in the present research work. In future work, the use of continuously evolving, synchronised photospheric magnetograms, in which the magnetic fields at different longitudes are observed simultaneously \citep{Upton2014a,Caplan_2025,Downs2025PSI}, together with accurate observations of the polar magnetic fields, could further reduce these discrepancies.

\begin{figure*}[htpb]
\begin{center}
  \vspace*{0.01\textwidth}
    \includegraphics[width=0.8\linewidth,trim=1 1 1 1, clip]{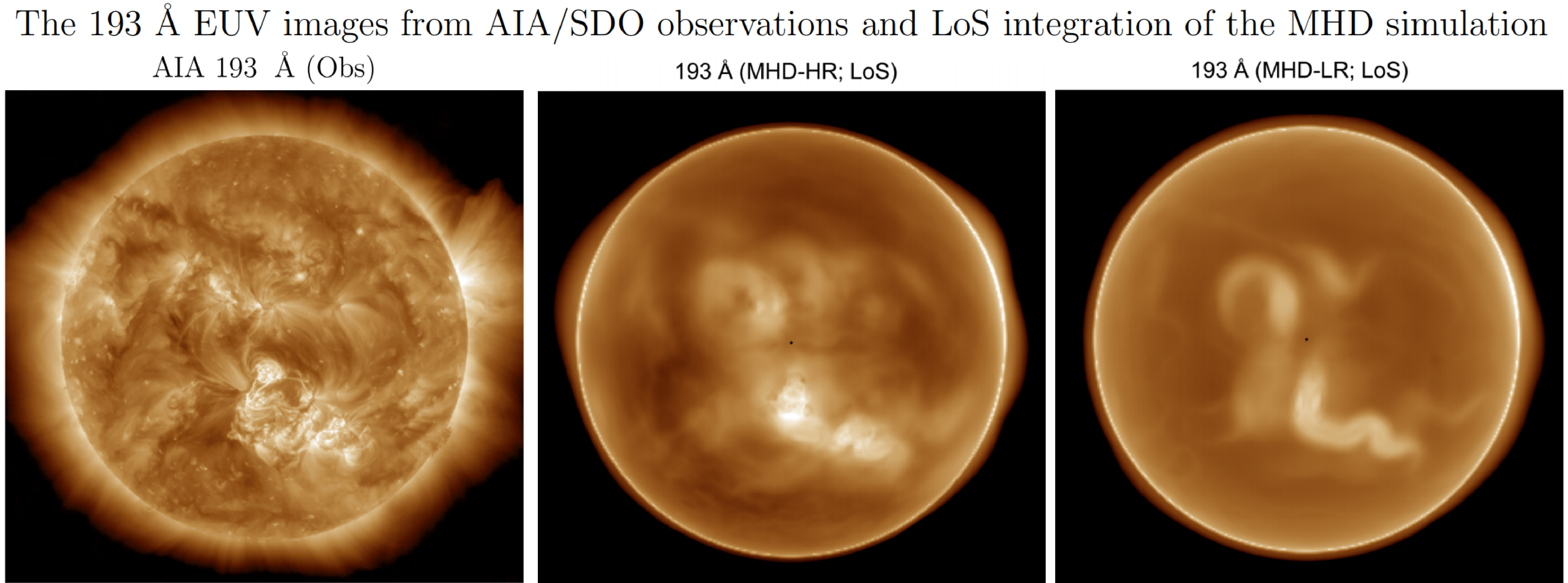}
\end{center}
\caption{The 193~{\AA} EUV images observed by AIA/SDO on June 15, 2012 (left) and derived from MHD-HR (middle) and MHD-LR (right) simulations.}\label{EUVSph}
\end{figure*}

In Fig.~\ref{EUVSph}, we present the full-disk 193~{\AA} EUV image observed by AIA/SDO on 15 June 2012 (left), together with synthetic images derived from MHD-HR (middle) and MHD-LR (right) simulations. These images are viewed from a heliocentric distance of 1~AU in the solar equatorial plane at a Carrington longitude of $87.5^{\circ}$. The MHD-HR simulation broadly reproduces the locations and relative intensities of the bright structures visible in the observation, particularly the clustered AR-related emission in the southern hemisphere. By contrast, filtering the magnetogram removes much of the small-scale magnetic structure, causing several distinct bright regions to merge into an overly smooth, simplified ribbon-like structure. Owing to the oversimplified empirical heating prescription, limited grid resolution, the unrealistic assumption of homogeneous flow-field conditions at the inner boundary, and the absence of a high-fidelity description of the complex thermal processes in the transition region and chromosphere, neither simulation reproduces the fine coronal-loop structures visible in the AIA/SDO image. Nevertheless, the coronal MHD simulation constrained by high-fidelity magnetogram yields a more structured distribution of EUV emission that is more consistent with the observations. 

Table~\ref{averagedopenflux} further lists the unsigned magnetic fluxes at selected heliocentric distances in both simulations, calculated following the procedure described in \cite{wang2026OpenFlux_ApJ}. The unsigned total magnetic flux which includes contributions from both open and closed magnetic fields, and the unsigned OMF at a heliocentric distance of $R_{\rm Se}$ are denoted by $\Phi^{R_{\rm Se}}_{\rm total}$ and $\Phi^{R_{\rm Se}}_{\rm open}$, respectively, while the area of the open-field regions at $R_{\rm Se}$ is denoted by $S^{R_{\rm Se}}_{\rm open}$.
The magnetic fluxes are normalised by $\Phi_{\rm 1~au}^{\rm ave}=7.98 \times 10^{14}~{\rm Wb}$, which is calculated using the WIND \textit{in situ} observations,
\begin{equation}\label{OMF_Interplanetary}
\Phi_{\rm 1~au}^{\rm ave}
=
\left| B_{r,\rm 1~au} \right|^{\rm ave}
\,\cdot\,4\pi\,(1~\mathrm{au})^{2}\,,    
\end{equation}
where $\left|B_{r,\rm 1~au}\right|^{\rm ave}$ denotes the average unsigned radial interplanetary magnetic field strength along the Sun-Earth line, derived from approximately 653 hourly averaged WIND observations \footnote{\url{https://cdaweb.gsfc.nasa.gov/index.html/}} \citep{king2005JGR} during the simulated CR, assuming a latitudinal invariance of the interplanetary magnetic field distribution \citep{Balogh1995,Smith1995,Lockwood2004}. The areas of the open-field regions at $r=3~R_s$ are normalised by the surface area of the corresponding sphere, $4\pi \cdot \left(3~R_s\right)^2$. 

As we know, a long-standing ``open flux problem" remains unresolved in solar and heliospheric physics \citep{Linker_2017,wang2026OpenFlux_ApJ}. Specifically, the OMF estimated from observation-based CH regions in synoptic LoS photospheric magnetograms is systematically smaller than that inferred from interplanetary in situ observations, and numerical extrapolations of OMF from observed magnetograms into the heliosphere generally underestimate the measured interplanetary OMF by a factor of two or more. The results in Table~\ref{averagedopenflux} demonstrate that, although the open-field regions at 3~$R_s$ are similar in both simulations, the MHD-HR simulation yields approximately $53.6\%$ more unsigned open magnetic flux and $56.3\%$ more unsigned total magnetic flux at 3~$R_s$, and $54.9\%$ more unsigned total magnetic flux at 0.1 au, compared with the MHD-LR simulation. 
This highlights the importance of using high-fidelity magnetograms for addressing the ``open flux problem" in coronal modelling.
\begin{table}
\centering
\caption{Simulated magnetic fluxes and areas of open-field regions}
\label{averagedopenflux}
\begin{tabular}{lllll}
\hline\noalign{\smallskip}
Case  & ${\rm S}^{3R_s}_{{\rm open}}$ & $\Phi^{3R_s}_{{\rm open}}$ & $\Phi^{3R_s}_{{\rm total}}$& $\Phi^{0.1\rm~au}_{{\rm total}}$\\
\noalign{\smallskip}\hline\noalign{\smallskip}
MHD-HR  & 0.86 &  1.29 & 1.36 & 1.10\\
MHD-LR  & 0.88 & 0.84 & 0.87 & 0.71 \\
\noalign{\smallskip}\hline
\end{tabular}
\end{table}

\subsection{Impact on the simulated coronal structures}\label{sec:CoronalStructures}
In this subsection, we further evaluate the impact of the generally filtered small-scale magnetic structures on the simulated coronal structures at selected meridional planes and heliocentric distances. We compare the coronal streamer structures obtained from the MHD-HR and MHD-LR simulations, and assess the agreement of the simulated coronal structures with coronal observations.

\begin{figure*}[htpb]
\begin{center}
  \vspace*{0.01\textwidth}
    \includegraphics[width=0.8\linewidth,trim=1 1 1 1, clip]{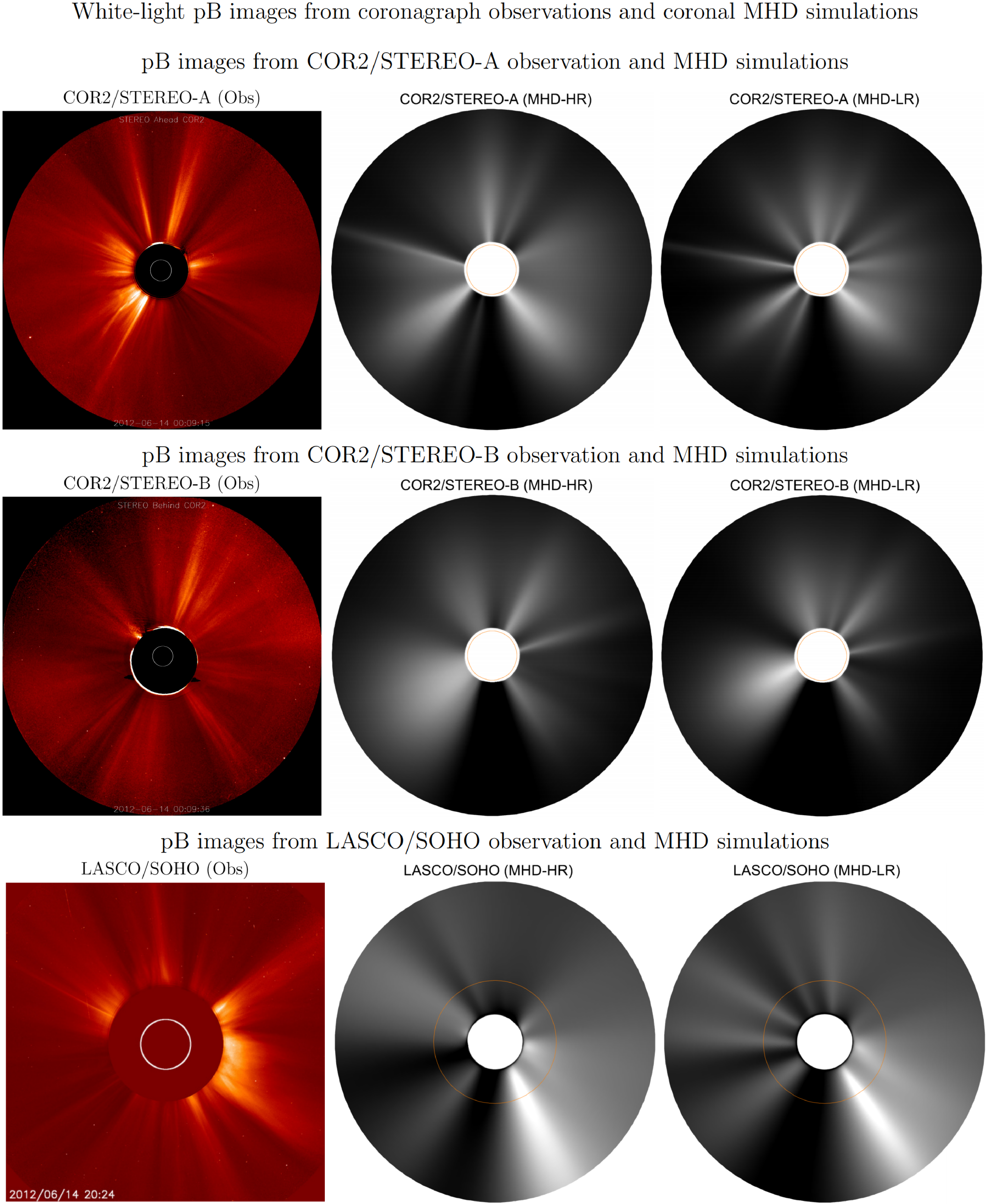}
\end{center}
\caption{The pB images observed from COR2/STEREO-A (left top) and COR2/STEREO-B (left middle), covering 2.5-15~$R_s$, and from LASCO-C2/SDO (left bottom), covering 2.3-6~$R_s$, are displayed alongside the corresponding pB images derived from MHD-HR (middle) and MHD-LR (right) simulations. The simulated pB images cover 2.5-15~$R_s$ for comparison with COR2/STEREO-A/B and 1.01-6~$R_s$ for comparison with LASCO-C2/SDO. The orange circles in the the simulated pB images indicate a heliocentric distance of 2.3~$R_s$.
}\label{pBSimulationandObs}
\end{figure*}

White-light pB images are commonly used to reveal various large-scale coronal structures. In pB images, bright features are generally associated with high-density coronal structures, such as streamers and pseudo streamers, whereas dark features are typically associated with low-density structures, such as CHs. In Fig.~\ref{pBSimulationandObs}, we present the pB images observed from the innermost coronagraph of the Sun Earth Connection Coronal and Heliospheric Investigation (SECCHI) instrument suite onboard the Solar Terrestrial Relations Observatory Ahead and Behind (STEREO-A/B; top-left and second-row-left panels) spacecrafts\footnote{\url{https://stereo-ssc.nascom.nasa.gov/browse/}} \citep{Howard2008} and the Large Angle and Spectrometric Coronagraph C2 (LASCO-C2) \citep{Brueckner1995} onboard the Solar and Heliospheric Observatory\footnote{\url{https://sdo.gsfc.nasa.gov/data/synoptic/}} (SOHO; bottom left), together with the corresponding simulated pB images \citep{Wang_COCONUT_DecE} spanning heliocentric distances from 2.5 to 15 $R_s$ (first and second rows) and from 1.01 to 6 $R_s$ (third row), derived from the MHD-HR (middle) and MHD-LR (right) simulations, respectively. At this moment, the Carrington longitudes of STEREO-A, STEREO-B, and SOHO are $205.8^{\circ}$, $331.6^{\circ}$, and $88.0^{\circ}$, respectively. The comparison shows that the MHD-HR and MHD-LR simulations produce broadly similar pB distributions, both of which basically reproduce the large-scale bright structures observed by COR2/STEREO-A and COR2/STEREO-B. Together with Fig.~3, the results suggest that the larger open-field region in the MHD-LR simulation alters the bright structures in the MHD-HR simulation around $20^{\circ}~{\rm N}$ and $60^{\circ}~{\rm S}$ on the west limb, separating them into two narrower bright structures divided by a dark lane. Furthermore, the MHD-HR simulation successfully reproduces the bright structure extending from approximately $5^{\circ}~{\rm N}$ to $80^{\circ}~{\rm N}$ and the narrow dark structure centred around $65^{\circ}~{\rm N}$ on the east limb in the LASCO-C2/SOHO observation, whereas the MHD-LR simulation produces a broader dark region over this latitude range.

\begin{figure*}[htpb]
\begin{center}
  \vspace*{0.01\textwidth}
    \includegraphics[width=0.8\linewidth,trim=1 1 1 1, clip]{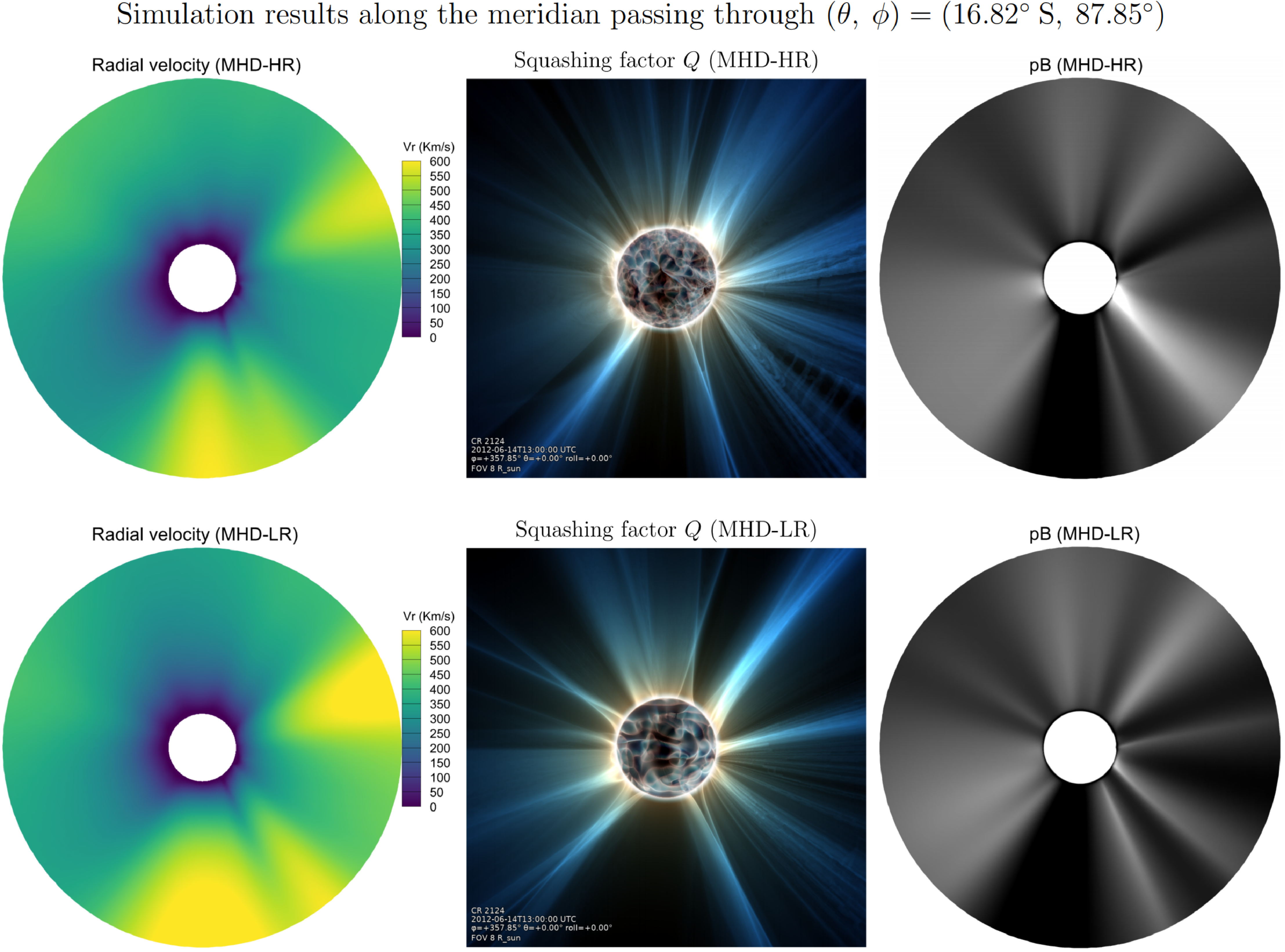}
\end{center}
\caption{The simulated radial velocity ranging from 1.01 to 5.7~$R_s$ (left; Km/S), simulated squashing factor distributions in a square of $11.4~R_s \times 11.4~R_s$ (middle) and  simulated pB images spanning from 1.01 to 5.7~$R_s$ (right) obtained from MHD-HR (top) and MHD-LR (bottom) simulations.
}\label{pBSimulation}
\end{figure*}

In Fig.~\ref{pBSimulation}, we further present the simulated radial velocity (left), pB images (right), and squashing factor distributions (middle) on the meridional plane perpendicular to the Sun-Earth line. The squashing factor panels were generated using the open-source package Qorona v0.4.0\footnote{\url{https://github.com/RayanDhib/Qorona}} \citep{Dhib2026Qorona,Qorona_v040}, which traces magnetic field lines and computes the perpendicular squashing factor, $Q_\perp$ \citep{Titov2007}. This quantity is large in quasi-separatrix layers, where field-line connectivity changes sharply, and diverges at separatrices. The renders thus reveal the magnetic topology underlying coronal fine structure, allowing a direct comparison of the fine scale magnetic structure in the two simulations. It reveals that regions with larger perpendicular squashing factors generally coincide with high-density, low-speed (HDLS) streamers. Compared with the MHD-LR simulation, the MHD-HR simulation produces substantially finer magnetic-topology structures and generally lower radial velocities. 
Table~\ref{averagedVrandNat3Rs} and \ref{averagedVrandNat0d1au} further list the area-averaged proton number density $N^{R_{\rm Se}}_{\rm Ave}$, radial velocity $V^{R_{\rm Se}}_{r,{\rm Ave}}$, and unsigned magnetic-field strength $\left|\mathbf{B}\right|^{R_{\rm Se}}_{\rm Ave}$ at a heliocentric distance of $R_{\rm Se}$ in the MHD-HR and MHD-LR simulations. Compared with the MHD-LR simulation, the MHD-HR simulation increases the averaged unsigned magnetic-field strength by $63.6\%$ and $66.2\%$, decreases the averaged radial velocity by $8.2\%$ and $8.6\%$, and increases the proton number density by $99.0\%$ and $100.0\%$ at 3~$R_s$ and 0.1~au, respectively.
\begin{table}
\centering
\caption{Area-averaged simulation results at 3~$R_s$}
\label{averagedVrandNat3Rs}
\begin{tabular}{llll}
\hline\noalign{\smallskip}
Case  & $N^{3R_s}_{\rm Ave} \left(\rm cm^{-3}\right)$ & $V^{3R_s}_{r,\rm Ave} \left(\rm Km\cdot S^{-1}\right)$ & $\left|\mathbf{B}\right|^{3R_s}_{\rm Ave} \left(\rm G\right)$ \\
\noalign{\smallskip}\hline\noalign{\smallskip}
MHD-HR  & $2.01\times10^7$ &  299.83 & 0.18 \\
MHD-LR  & $1.01\times10^7$ &  326.61 & 0.11  \\
\noalign{\smallskip}\hline
\end{tabular}
\end{table}
\begin{table}
\centering
\caption{Area-averaged simulation results at 0.1~au}
\label{averagedVrandNat0d1au}
\begin{tabular}{llll}
\hline\noalign{\smallskip}
Case  & $N^{3R_s}_{\rm Ave} \left(\rm cm^{-3}\right)$ & $V^{3R_s}_{r,\rm Ave} \left(\rm Km\cdot S^{-1}\right)$ & $\left|\mathbf{B}\right|^{3R_s}_{\rm Ave} \left(\rm nT\right)$ \\
\noalign{\smallskip}\hline\noalign{\smallskip}
MHD-HR  & $2.90\times10^5$ &  438.43 & 344 \\
MHD-LR  & $1.45\times10^5$ &  479.73 & 207 \\
\noalign{\smallskip}\hline
\end{tabular}
\end{table}

\begin{figure*}[htpb]
\begin{center}
  \vspace*{0.01\textwidth}
    \includegraphics[width=0.8\linewidth,trim=1 1 1 1, clip]{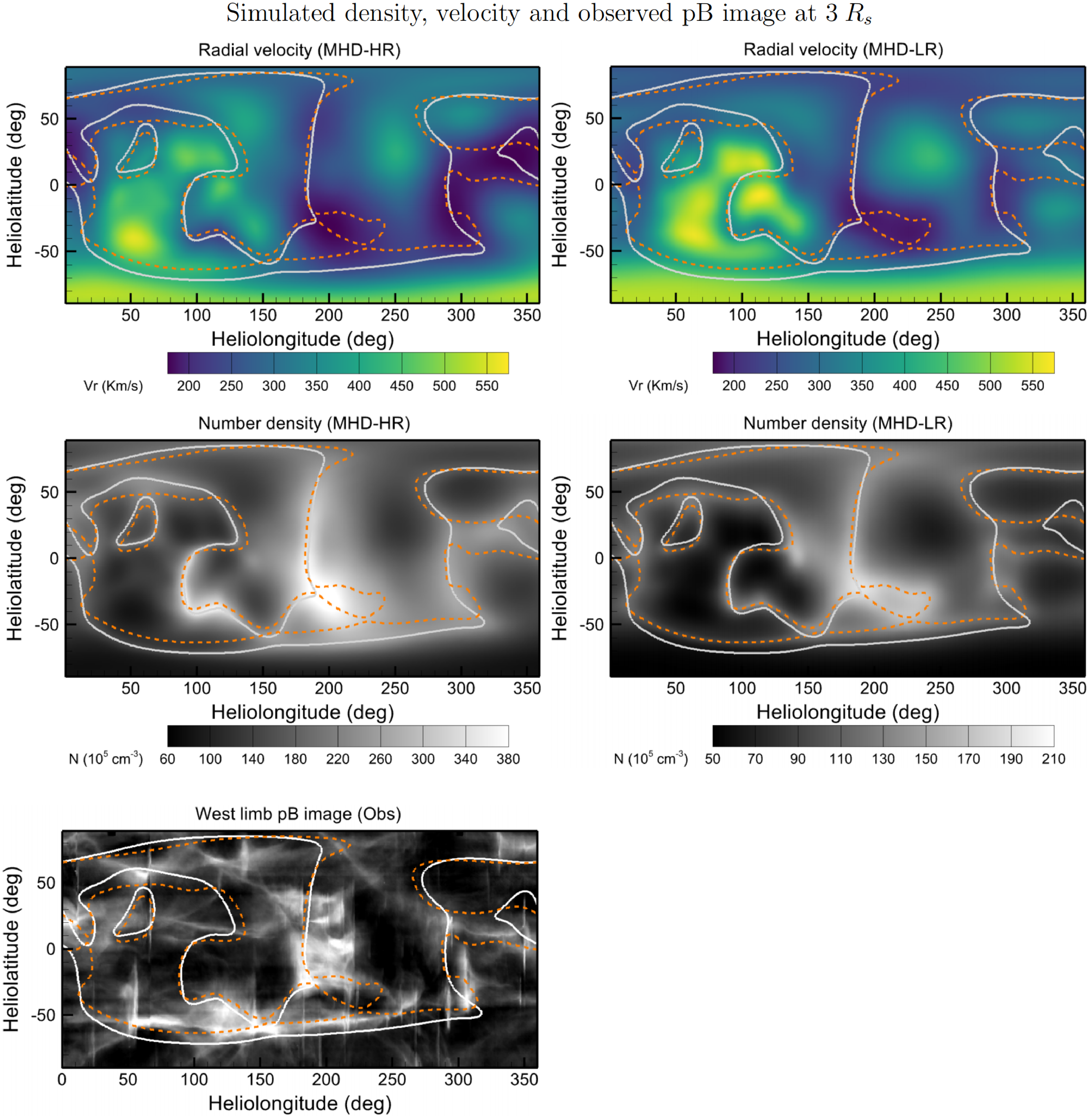}
\end{center}
\caption{The simulated radial velocity (first row) and proton number density (second row) from the MHD-HR (left) and MHD-LR (right) simulations, respectively, together with the synoptic maps of west-limb white-light pB observation (third row) from the SOHO instrument LASCO C2 at 3 $R_s$ for CR 2124. The solid gray line and dashed orange line represent the corresponding magnetic neutral lines (MNLs) derived from the coronal simulations constrained by the high-fidelity and filtered magnetograms, respectively.}\label{pBPannel}
\end{figure*}

In Fig.~\ref{pBPannel}, we display the simulated radial velocity (first row) and proton number density distributions (second row) at 3~$R_s$ from MHD-HR (left) and MHD-LR (right) simulations. We also present the synoptic map of the west-limb pB observations (third row) at 3~$R_s$ (left), obtained from the Large Angle and Spectrometric Coronagraph C2 (LASCO-C2) \citep{Brueckner1995} onboard the Solar and Heliospheric Observatory (SOHO)\footnote{\url{https://sdo.gsfc.nasa.gov/data/synoptic/}} for CR 2124. The results demonstrate that both the MHD-HR and MHD-LR simulations reproduce the principal high-density structures revealed by the pB observations. The simulated HDLS streamer structures closely follow the corresponding magnetic neutral lines. Moreover, the MHD-HR simulation successfully reproduces the observed bright pB structure located near $15^{\circ}~{\rm N}$ between Carrington longitudes of $300^{\circ}$ and $360^{\circ}$, whereas this feature is only weakly reproduced in the MHD-LR simulation. 

\section{Concluding remarks}\label{sec:Conclusion}
High-fidelity photospheric magnetograms are essential for constraining coronal MHD models. However, owing to limitations in grid resolution, numerical stability, and computational efficiency, the magnetograms used to prescribe the inner-boundary magnetic field of global coronal MHD simulations are commonly smoothed or reconstructed using truncated spherical-harmonic expansions. Such preprocessing removes a substantial fraction of the small-scale magnetic structures and may limit the ability of the models to reproduce fine-scale coronal features.
Recently, we have improved the numerical stability of COCONUT, a highly efficient implicit MHD coronal model, using an energy decomposition strategy combined with a HLL Riemann solver, in which an extra dissipation term is introduced into the energy equation. This makes COCONUT numerically stable for most coronal simulations starting from the low corona, where the magnetic field strength does not significantly exceed 100 G. In this paper, we incorporate a simplified TR module into COCONUT and locally refine the grid mesh around the selected AR to enable coronal simulations constrained by high-fidelity magnetograms that preserve most small-scale magnetic structures, with magnetic-field strengths exceeding 1000~G near the AR.

Using the COCONUT model developed in this study, we perform coronal simulations constrained by the high-fidelity and filtered magnetograms to evaluate the effects of the small-scale magnetic structures that are typically removed in MHD coronal simulations. The high-fidelity magnetogram is synthesised from central-meridian strips extracted from a series of hourly updated full-disc HMI magnetograms centred on 14 June 2012. The high-fidelity magnetogram is synthesised by updating observations of HMI magnetogram centred on 2012 June 14 from the HMI Polar-Filled synoptic maps. It is preprocessed using only Gaussian smoothing, thereby retaining most of the fine-scale magnetic structures and a maximum magnetic-field strength of approximately 1500~G. By contrast, the filtered magnetogram is reconstructed using a 50th-order PF solver with filtered spherical harmonic expansion \citep{MCCLARREN20105597,Murteira2025,wang2026OpenFlux_ApJ}, which filters out much of the small-scale structures, reducing the maximum magnetic-field strength to approximately 80~G.

Performed on the same grid and using the same governing equations, the MHD-HR simulation, constrained by the high-fidelity magnetogram, produces approximately $50\%$ more unsigned total magnetic flux and OMF at both 3~$R_s$ and 0.1~au than the MHD-LR simulation constrained by the filtered magnetogram. This highlights the importance of using high-fidelity magnetograms for addressing the long-standing ``open flux problem" in coronal modelling. Meanwhile, compared with the MHD-LR simulation, the retention of many more small-scale closed-field structures in the MHD-HR simulation increases the mean plasma density by $99.0\%$ and $100.0\%$ and decreases the mean radial velocity by $8.2\%$ and $8.6\%$ at 3~$R_s$ and 0.1~au, respectively. These differences highlight the need to incorporate more self-consistent thermodynamic mechanics into coronal simulations constrained by high-fidelity magnetograms, as the increased magnetic complexity, which play a key role in the thermodynamic processes in low corona, can have a non-negligible impact on the resulting magnetic-field and plasma-flow distributions.
Furthermore, the OMF obtained from the MHD-HR simulation at 0.1~au is approximately 1.1 times the value inferred from WIND \textit{in situ} observations. In a previous coronal simulation conducted during two CRs near solar maximum, constrained by filtered magnetograms, and performed on a level-six grid, the OMF obtained at 0.1~AU was only approximately $50\%$ of the value inferred from \textit{in situ} observations \citep{wang2026OpenFlux_ApJ}. The increased OMF obtained in the present study can be attributed primarily to the use of a high-fidelity magnetogram and higher grid resolution. 

Additionally, the MHD-HR generally reproduces the large-scale coronal structures observed in pB images from COR2/STEREO-A, COR2/STEREO-B, and LASCO-C2/SOHO, as well as the low- and mid-latitude structures observed in the 193~{\AA} EUV channel of AIA/SDO. 
Although the synthetic pB images from the MHD-LR simulation contain more extensive dark regions than those from the MHD-HR simulation, their principal bright structures are broadly consistent. By contrast, the synthetic EUV images from the two simulations exhibit substantial differences, with the MHD-HR results displaying more intricate dark structures and higher-contrast bright features, thereby showing better agreement with the observations. The simulation results also show that the MHD-HR simulation produces a finer-scale distribution of the perpendicular squashing factor, indicating that it resolves substantially more intricate magnetic-topology structures. 
Moreover, a comparison among the simulated open-field distributions, synthetic EUV images obtained by radial and LoS integration, and corresponding AIA/SDO observations shows that the open-field polar regions appear dark in the radially integrated images but may appear as bright structures in Earth-viewed LoS images. However, such bright emission is typically associated with closed-field regions, demonstrating the ambiguity introduced when the solar poles are observed from low latitudes. This finding underscores the importance of solar polar-orbiting missions \citep{Deng2025ChJSS}, which can provide reliable and accurate observations of the polar regions and significantly improve full-Sun coverage observations beyond what is achievable from conventional ecliptic-plane viewpoints.

Building on advances of COCONUT in computational efficiency and numerical stability, together with the capability to perform coronal simulations directly constrained by high-fidelity photospheric magnetograms, our future work will focus on incorporating more self-consistent, physics-based treatments of coronal heating and solar-wind acceleration mechanisms \citep{Cranmer_2010,Schleich2023,VanDoorsselaere2025}. We will further simulate the CME event \citep{WangWS2019} based on the MHD-HR and MHD-LR background coronal simulations presented in this research work to evaluate the effects of the small-scale magnetic-field structures, which are typically removed in coronal MHD simulations, on CME evolutions and propagations. We will subsequently extend the CME simulation to a time-evolving coronal background, and finally realise self-consistent, data-driven CME initiation and eruption simulations \cite[e.g.][]{Jiang2021fRONTIER,Jiang2023} in a time-evolving Sun-to-Earth model \citep{wang2026coconutpracticaltimeevolvingsuntoearth}, thereby advancing the model towards practical applications in daily space-weather forecasting. Additionally, we plan to incorporate the advective flux transport (AFT) model \citep{Upton2014a} with observations from the Solar Orbiter Polarimetric and Helioseismic Imager \citep{Loeschl2024} to produce more realistic synchronised magnetograms \citep{Upton2014a,Caplan_2025,Downs2025PSI}. These magnetograms will be used to drive our MHD coronal model to reconstruct the continuously evolving full-solar-cycle coronal evolution,  enabling us to investigate the long-term evolution of the coronal and heliospheric current sheets and to assess how the reversal of the solar polar magnetic field affects the evolution of the corona and solar wind.
We will also incorporate far-side data reconstructed using an artificial intelligence-based model to improve the inner boundary conditions and modelling results for regions near the far side \citep{Jeong2020Solar,Jeong2026Revisiting}.

\begin{acknowledgements}
This project has received funding from the European Research Council Executive Agency (ERCEA) under the ERC-AdG agreement No. 101141362 (Open SESAME). 
These results were also obtained in the framework of the projects FA9550-18-1-0093 (AFOSR), C16/24/010  (C1 project Internal Funds KU Leuven), G0B5823N and G002523N (WEAVE) (FWO-Vlaanderen), and 4000145223 (SIDC Data Exploitation (SIDEX), ESA Prodex).
This work is also supported by the BK21 FOUR program of the Graduate School, Kyung Hee University (GSX-20253142). The resources and services used in this work were provided by the VSC (Flemish Supercomputer Centre), funded by the Research Foundation – Flanders (FWO) and the Flemish Government. This work utilises data obtained by the Global Oscillation Network Group (GONG) program, managed by the National Solar Observatory and operated by AURA, Inc., under a cooperative agreement with the National Science Foundation. The data were acquired by instruments operated by the Big Bear Solar Observatory, High Altitude Observatory, Learmonth Solar Observatory, Udaipur Solar Observatory, Instituto de Astrof{\'i}sica de Canarias, and Cerro Tololo Inter-American Observatory. The authors also acknowledge the use of the STEREO/SECCHI data produced by a consortium of the NRL (US), LMSAL (US), NASA/GSFC (US), RAL (UK), UBHAM (UK), MPS (Germany), CSL (Belgium), IOTA (France), and IAS (France). 
\end{acknowledgements}

\bibliographystyle{aa}
\bibliography{SIPandCOCONUT}

\end{document}